\documentclass[aps, reprint, notitlepage,  superscriptaddress, pra]{revtex4-1}
\usepackage{blindtext}
\usepackage{graphicx}
\usepackage{xcolor}
\usepackage{amsmath}
\usepackage{amsthm}
\usepackage{amssymb}
\usepackage{amsfonts}
\usepackage{xr-hyper}
\usepackage{hyperref}
\hypersetup{
    colorlinks,
    linkcolor={blue},
    citecolor={blue},
    urlcolor={blue}
}
\usepackage{siunitx}
\usepackage{bm}
\usepackage{ulem,xpatch}
\usepackage{lineno}
\usepackage{indentfirst}

\newcommand{\dd}{\mathrm{d}}

\newcommand{\inn}{\mathrm{in}}
\newcommand{\outt}{\mathrm{out}}

\newcommand{\imu}{\text{\rm i}}
\newcommand{\expu}{\text{\rm e}}
\newcommand{\Og}{\Omega}
\newcommand{\Om}{\Omega_m}

\newcommand{\Gmeas}{\Gamma_\text{meas}}
\newcommand{\Gqba}{\Gamma_\text{qba}}

\newcommand{\Gtot}{\Gamma_\text{tot}}
\newcommand{\etad}{\eta_\mathrm{d}}

\newcommand{\gm}{\gamma_m}

\newcommand{\affil}{Photonics Laboratory, ETH Zürich, CH-8093 Zürich, Switzerland}
\newcommand{\affilQC}{Quantum Center, ETH Zürich, CH-8093 Zürich, Switzerland}
\newcommand{\affilIQOQI}{Institute for Quantum Optics and Quantum Information of the Austrian Academy of Sciences, A-6020, Innsbruck, Austria}
\newcommand{\affilITP}{Institute for Theoretical Physics, University of Innsbruck, A-6020 Innsbruck, Austria}
\newcommand{\equalcontribution}{These authors contributed equally to this work.}

\begin{document}

\title{Ponderomotive squeezing of light by a levitated nanoparticle in free space}
\author{Andrei Militaru}
\altaffiliation{\equalcontribution}
\affiliation{\affil}
\author{Massimiliano Rossi}
\altaffiliation{\equalcontribution}
\affiliation{\affil}
\author{Felix Tebbenjohanns}
\altaffiliation{present address: Department of Physics, Humboldt-Universität zu Berlin, 10099 Berlin, Germany}
\affiliation{\affil}
\author{Oriol Romero-Isart}
\affiliation{\affilIQOQI}
\affiliation{\affilITP}
\author{Martin Frimmer}
\affiliation{\affil}
\author{Lukas Novotny}
\affiliation{\affil}
\affiliation{\affilQC}

\begin{abstract}
  A mechanically compliant element can be set into motion by the interaction with light.
  In turn, this light-driven motion can give rise to ponderomotive correlations in the electromagnetic field.
  In optomechanical systems, cavities are often employed to enhance these correlations up to the point where they generate quantum squeezing of light.
  In free-space scenarios, where no cavity is used, observation of squeezing remains possible but challenging due to the weakness of the interaction, and has not been reported so far.
  Here, we measure the ponderomotively squeezed state of light scattered by a nanoparticle levitated in a free-space optical tweezer.
  We observe a reduction of the optical fluctuations by up to $25$~\% below the vacuum level, in a bandwidth of about $15$~kHz.
  Our results are well explained by a linearized dipole interaction between the nanoparticle and the electromagnetic continuum.
  These ponderomotive correlations open the door to quantum-enhanced sensing and metrology with levitated systems, such as force measurements below the standard quantum limit.
\end{abstract}

\maketitle

Cavity-enhanced light-matter interaction is a central paradigm in condensed-matter physics \cite{garcia-vidal_manipulating_2021}, especially in the fields of cavity and circuit quantum electrodynamics \cite{berman_cavity_1994, blais_circuit_2021}.
More recently, researchers in cavity optomechanics have employed similar techniques in order to measure and control the motion of solid-state systems, from nanomechanical resonators to kilogram-scale mirrors \cite{aspelmeyer_cavity_2014-1}.

Electromagnetic resonators come also with drawbacks, such as bandwidth limitations and reduced coupling efficiencies due to mode mismatching \cite{roy_colloquium_2017}.
To circumvent these problems, there exist alternative coupling schemes which make use of travelling electromagnetic fields either in waveguides or directly in free space, rather than in cavities.
In the context of optomechanics, these schemes have been studied with Brillouin and Raman scattering from bulk acoustic waves \cite{renninger_bulk_2018, otterstrom_optomechanical_2018} and in levitodynamics with optical tweezers \cite{ashkin_optical_1976, libbrecht_toward_2004, gonzalez-ballestero_levitodynamics_2021}.

The latter scenario is an example of a free-space system: here, an optical trap is formed for a dielectric nanoparticle by tightly focusing an intense laser field.
The nanoparticle imprints a position-dependent phase to the scattered laser photons.
Interferometric techniques allow one to retrieve this phase, effectively realizing a displacement measurement.
At the same time, the nanoparticle recoils after a photon scattering event, which occurs at a stochastic rate.
This is a form of quantum backaction and generates fluctuations in the nanoparticle position.
Recent experimental advances have made possible to access a regime in which the quantum backaction is the dominant source of position fluctuations, which are efficiently recorded in phase measurements \cite{jain_direct_2016, tebbenjohanns_motional_2020}.
These advances enabled measurement-based ground-state cooling of the motional state of nanoparticles in free-space levitodynamics \cite{magrini_real-time_2021, tebbenjohanns_quantum_2021}.

In addition to ground-state cooling, this quantum regime of measurement enables the generation of quantum correlations in the mode of the scattered light.
The nanoparticle motion correlates the optical amplitude and phase quadratures, which are responsible for the quantum backaction and measurement imprecision, respectively.
If strong enough, these correlations may lead to a reduction of the fluctuations of an optical quadrature below the level of vacuum fluctuations, a phenomenon known as ponderomotive squeezing \cite{fabre_quantum-noise_1994, mancini_quantum_1994}.
In optomechanics, this quantum squeezing has been observed with ultracold atoms \cite{brooks_non-classical_2012}, optomechanical photonic crystals \cite{safavi-naeini_squeezed_2013}, membrane resonators \cite{purdy_strong_2013,  nielsen_multimode_2017, chen_entanglement_2020}, and crystalline cantilevers \cite{aggarwal_room-temperature_2020}.
All these experiments are based on a cavity-enhanced optomechanical interaction.
To date, no observations of ponderomotive squeezing in free-space optomechanical systems have been reported.

In this work, we measure squeezing by 25\% below the vacuum noise in the light scattered by a levitated nanoparticle in free space.
We fully reconstruct the state of the squeezed optical mode by homodyne tomography \cite{vogel_determination_1989, smithey_measurement_1993, lvovsky_quantum_2001}.
Furthermore, we explain our experiments with quantum optics theory, which assumes a linearized dipole interaction between the nanoparticle motion and the electromagnetic field.

\begin{figure}
  \includegraphics{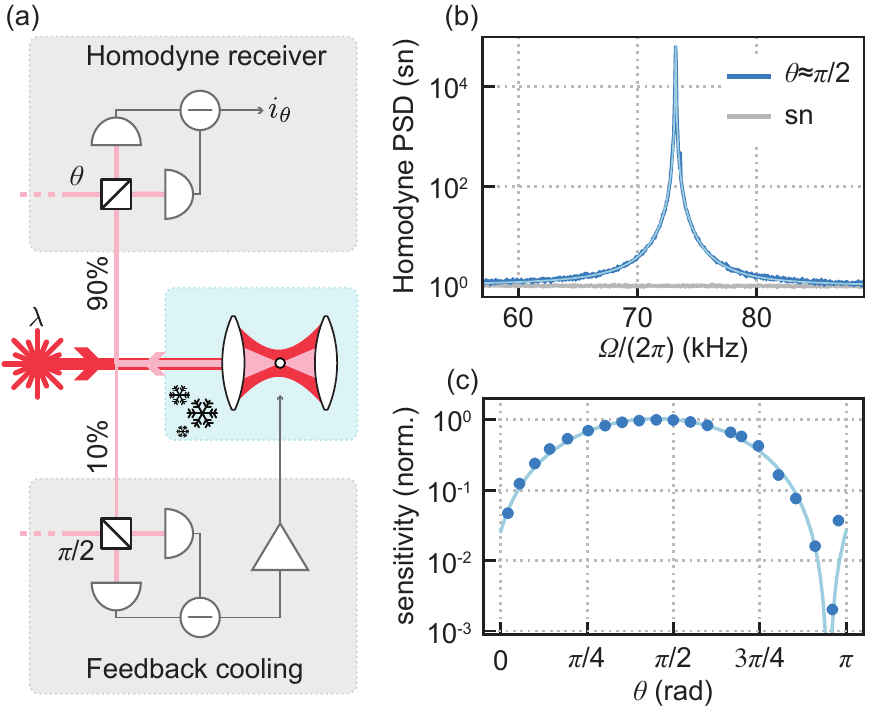}
  \caption{Free-space levitated optomechanics. (a) Experimental setup. (b) Power spectral density (PSD) of a phase quadrature measurement, normalized to the shot-noise (sn) background. (c) Sensitivity to mechanical motion, normalized to its maximum. The sensitivity is measured as the detector response to an off-resonant sinusoidal force acting on the particle. The light blue line is a sinusoidal fit.}
  \label{fig:exp}
\end{figure}

Our experimental setup is shown in Fig.~\ref{fig:exp}(a) and consists of a spherical silica nanoparticle of 100~nm diameter trapped in the focus of an optical tweezer.
The laser (wavelength 1550~nm, power 1.2~W) is linearly polarized along the $x$ direction and propagates along the direction $z$.
In the following, we will only consider the motion along this longitudinal direction ($z$ axis).
We form the optical tweezer by strongly focusing the laser by an aspheric lens which is located inside a $4$~K cryostat.
More details can be found in Ref.~\cite{tebbenjohanns_quantum_2021}.
Once cooled down, the consequent cryogenic pumping mechanism evacuates the volume around the optical trap to a pressure below $10^{-9}$~mbar.
At this pressure, the quantum backaction from the photon scattering dominates over the motional decoherence induced by collisions with the surrounding gas molecules \cite{jain_direct_2016}.
We perform homodyne detection to monitor the field scattered by the nanoparticle, characterized by the amplitude and phase quadratures $X_\outt$ and $Y_\outt$, respectively.
To do so, we overlap the scattered light with a strong coherent local oscillator beam (LO) with relative phase $\theta$.
We optimize the LO Gaussian transverse profile to match the one of the scattered light in the backward direction, such that the detection efficiency $\eta_\mathrm{d}$ of the longitudinal motion is maximized \cite{tebbenjohanns_optimal_2019}.
We model the losses and finite detection efficiency with a fictitious beamsplitter of transmissivity $\eta_d$ in front of the detector \cite{yuen_optical_1980}.
The homodyne photocurrent becomes $i_\theta = \sqrt{\eta_\mathrm{d}} X_\text{out}^\theta + \sqrt{1-\eta_\mathrm{d}} X_\nu$, where $X_\nu$ is the amplitude quadrature of an uncorrelated field in the vacuum state entering from the dark port of the fictitious beamsplitter. In the equation for the photocurrent, we have introduced the rotated quadrature of the scattered field
\begin{eqnarray}
  \label{eq:7}
  X^\theta_\text{out} = \cos(\theta) X_\text{out} + \sin(\theta) Y_\text{out}.
\end{eqnarray}
In the experiment, we split the scattered field in two parts, as shown in Fig.~\ref{fig:exp}(a).
A small fraction (10\%) is used to perform a homodyne phase measurement to feedback-cool the particle motion \cite{tebbenjohanns_quantum_2021}.
The remaining part (90\%) is sent to a different, out-of-loop homodyne receiver.
This is the main detector of our experiments and we use it to measure ponderomotive squeezing.
For this detector, we stabilize its LO phase, $\theta$, to any value in the range $[0, \pi]$ to measure the corresponding optical quadrature.

In Fig.~\ref{fig:exp}(b), we show the power spectral density (PSD) of the homodyne photocurrent, $S_{ii}^\theta$, for a phase quadrature measurement ($\theta\approx\pi/2$) \footnote{We adopt the following definition $S_{ii}^\theta(\Omega)~=~1/(2\pi) \int_\mathbb{R}\mathrm{d}\tau\ e^{i\Omega\tau} \langle\overline{i_\theta(t)i_\theta(t+\tau)}\rangle$, where the overline indicates a symmetrized quantity.}.
The flat noise floor in the spectrum arises from the vacuum noise of the probing light field.
On top of this background stands a Lorentzian peak, which represents the nanoparticle motion.
By fitting the PSD with a Lorentzian function, we extract the mechanical resonance frequency $\Om/(2\pi)=73.25$~kHz and the damping rate $\gm/(2\pi)=40$~Hz.
This damping rate results from the mild feedback cooling exerted on the nanoparticle.

We now measure the PSD as we change the angle $\theta$ in the range $[0, \pi]$.
This reduces the sensitivity of our measurements of the particle motion, which is solely contained in the phase quadrature $Y_\outt$.
To characterize this sensitivity, we exert an off-resonant (\SI{90}{kHz}) sinusoidal force on the nanoparticle.
We exert this force electrically, which is possible thanks to the net charge carried by the nanoparticle  \cite{frimmer_controlling_2017, tebbenjohanns_quantum_2021}.
The driven motion appears in the homodyne photocurrent as a sinusoidal oscillation at \SI{90}{kHz}.
We record its amplitude for different angles $\theta$, as shown in Fig.~\ref{fig:exp}(c).
The maximum response occurs around the phase quadrature, at angle $\pi/2$, whereas the minimum response is shifted from $\pi$ by $\sim0.05\pi$.
This deviation from $\pi$ is caused by a additional, weak reflection of the tweezer light copropagating backward with the scattered light towards the homodyne detector.

\begin{figure}[b]
  \centering
  \includegraphics{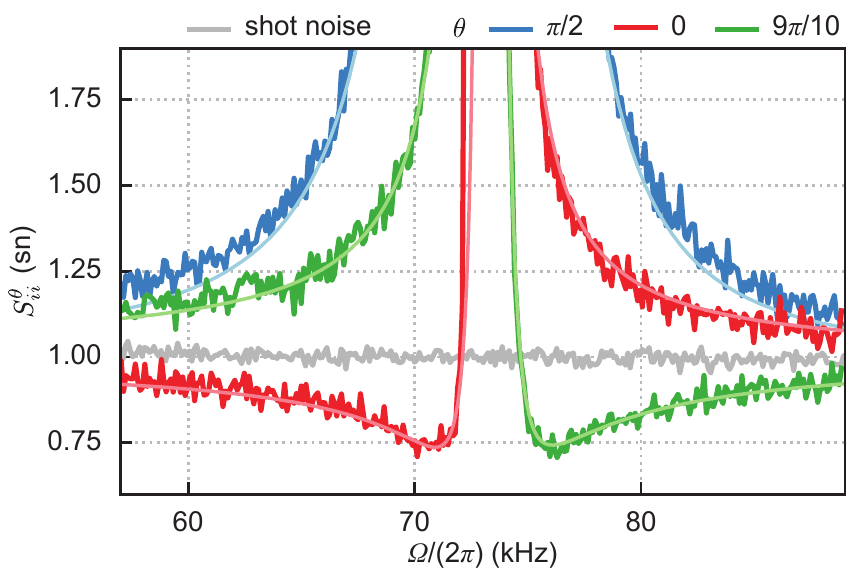}
  \caption{Ponderomotive squeezing.
    Enlarged view of three different PSDs, at the phase quadrature (blue) and close to the amplitude one (red and green).
    The solid light-coloured lines are the results of the fits to Eq.~\eqref{eq:psd_theory}.
    The gray trace is the measured shot noise.
    We have subtracted from the PSDs a small classical noise contribution originating from the LO (see Supplemental).
}
  \label{fig:squeezing_1}
\end{figure}

In Fig.~\ref{fig:squeezing_1}, we compare two PSDs acquired close to the amplitude quadrature, at $\theta\approx 0$ (red) and at $\theta\approx0.9\pi$ (green), with one at the phase quadrature $\theta\approx\pi/2$ (blue).
The former two show an asymmetric Fano lineshape rather than a Lorentzian one, suggesting interference between a broadband background, generated by both optical quadratures, and a resonant process, generated by the mechanical motion driven by the optical amplitude quadrature.
The measured spectral noise lies below the shot noise within about $15$~kHz bandwidth, with a maximum noise reduction of $25$\%.
This is the manifestation of ponderomotive squeezing of the scattered light field \cite{fabre_quantum-noise_1994, mancini_quantum_1994}.

We now perform homodyne tomography on the backscattered field, in order to fully reconstruct the squeezed states.
For each angle $\theta$, we extract realizations of a temporal mode from the homodyne photocurrent according to
\begin{eqnarray}
  \label{eq:1}
  r^\theta_{\Og} = \int_{-T/2}^{T/2}\dd t\,\expu^{\imu \Og t}\,i_\theta(t),
\end{eqnarray}
where $T\approx8~\mathrm{ms}$ is chosen to be larger than the correlation time $1/\gamma_m$ in order to consider statistically independent realizations.
In the spectral domain, this temporal mode corresponds to a frequency bin centered at $\Omega$ and with a width of $1/T=121~\mathrm{Hz}$.
We collect an ensemble of $\sim10^4$ realizations for both the real and imaginary part of $r^\theta_\Og$, then we compute their histograms.
We repeat this procedure for different angles $\theta$.
The histograms correspond to marginals of the Wigner quasiprobability distribution $\mathcal{W}(X,Y)$ along the angle $\theta$ with respect to the $X$ axis \cite{breitenbach_measurement_1997, smithey_measurement_1993, lvovsky_continuous-variable_2009}. 
To reconstruct the function $\mathcal{W}(X,Y)$, we apply the inverse Radon transform to the set of histograms (see Supplemental).
In Fig.~\ref{fig:tomography}, we show the experimentally reconstructed Wigner functions for two modes centered at $\Og/(2\pi)=$~\SI{70.1}{kHz} in (a) and \SI{77.1}{kHz} in (b), for which we have the strongest correlations.
\begin{figure}[t]
  \centering
   \includegraphics{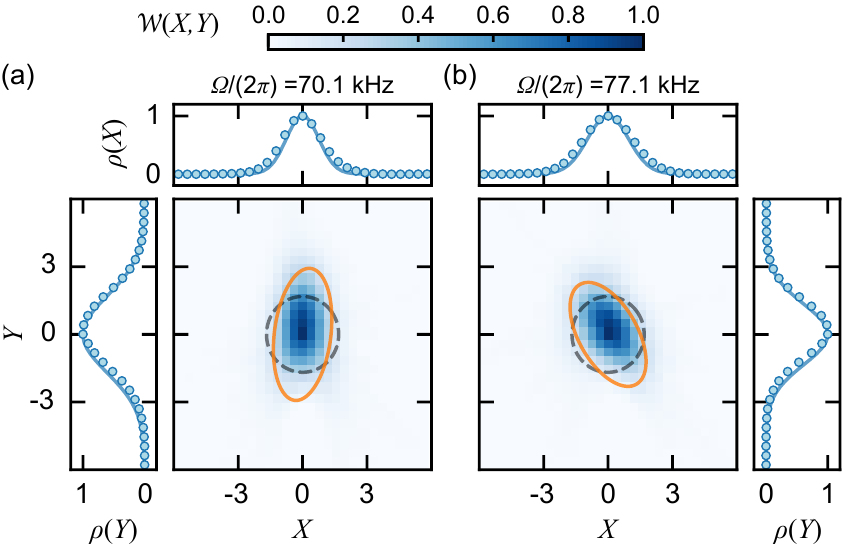}
  \caption{Homodyne tomography of ponderomotive squeezed light.
    (a), (b) Reconstructed Wigner functions of modes centered at, respectively, $\Omega/(2\pi)=70.1$~kHz and  $77.1$~kHz.
    The solid orange (dashed gray) line indicates the covariance ellipse with two standard deviations (s.d.) of the squeezed (vacuum) state.
    Above and aside the Wigner functions we show the marginals distributions, $\rho$, for $X$ and $Y$, respectively.
    Blue dots are data, the light blue lines are theoretical predictions (see Supplemental).
    The optical quadratures are normalized such that the vacuum noise s.d. is $1/\sqrt{2}$.
    }
  \label{fig:tomography}
\end{figure}
Since the optical states are Gaussian, the Wigner function is fully determined by the quadratures' means and covariance matrix. The latter can be simply estimated from three PSDs at different angles (see Supplemental).
We apply this idea to estimate the covariance ellipses shown in Fig.~\ref{fig:tomography} .
The covariance ellipse from the scattered light (solid orange line) is narrower than the one of the vacuum state (dashed gray line) along some directions, a distinctive fingerprint of squeezing.
We also notice that the angle of squeezing changes with the frequency $\Omega$ of the temporal mode.

Light squeezing via the motion of a mechanical system induced by quantum backaction is known as ponderomotive squeezing, a phenomenon that has been recently observed in optomechanics \cite{brooks_non-classical_2012,safavi-naeini_squeezed_2013,purdy_strong_2013,nielsen_multimode_2017, chen_entanglement_2020,aggarwal_room-temperature_2020}.
Therein, the squeezed optical mode is determined by the cavity resonance driven by the laser and interacting with the mechanical oscillator.
In contrast, in our case the nanoparticle simultaneously interacts with all the modes of the electromagnetic continuum due to the absence of a cavity.
We analyze our experiments with a theoretical framework, which assumes a linearized dipolar light-matter interaction \cite{chang_cavity_2010, romero-isart_optically_2011, rodenburg_quantum_2016, gonzalez-ballestero_theory_2019, magrini_real-time_2021}.
The longitudinal nanoparticle motion interacts with a set of plane waves of the electromagnetic field (similarly for the motion along other directions).
This combination defines a distinct optical mode, the amplitude and phase quadratures of which we label $X_\inn$ and $Y_\inn$ (see Supplemental).
The spatial distribution of this mode, which we term {\it interacting mode}, has been first derived in Ref.~\cite{tebbenjohanns_optimal_2019}, where it played the role of information density patterns in an optimal position measurement of a dipolar scatterer.

With the use of the interacting mode and in the interaction picture with respect to the free field, the interaction Hamiltonian of our system is $H_\mathrm{int} = -\sqrt{4\Gqba}X_\mathrm{in}\,q$, where $\Gqba$ is a rate characterizing the interaction strength, and $q$ is the position operator normalized to $\sqrt{\hbar/(m\Omega_m)}$, with $m$ the nanoparticle mass. After including the harmonic potential generated by the optical tweezer, the Heisenberg equation of motion for $q$ becomes
\begin{eqnarray}
  \label{eq:main_eom_q}
  \ddot{q} +\Og_m^2 q + \gamma_m \dot{q} = \Og_m\left(\xi(t) + \sqrt{4\Gqba}\ X_\inn(t)\right)
\end{eqnarray}
where we introduced the damping rate $\gamma_m$ which accounts for the surrounding gas and for the feedback cooling.
The term $\xi$ is a white thermal force that satisfies the correlation $\langle\overline{\xi(t)\xi(t')}\rangle=2\gm\left(\overline{n}+1/2\right)\delta(t-t')$, where $\overline{n}$ is the average phonon occupancy \cite{giovannetti_phase-noise_2001, symmetrization_footnote}.
The last term in Eq.~\eqref{eq:main_eom_q} is the quantum backaction exerted on the particle by the interacting mode.
At the same time, the nanoparticle motion affects the quadratures ($X_\inn$, $Y_\inn$) of the interacting modes. The resulting quadratures ($X_\outt$, $Y_\outt$), labelled output, are derived from the input-output relation  \cite{gardiner_input_1985}
\begin{subequations}\label{eq: input-output}
\begin{align}
  X_\outt(t) &= X_\inn(t),\label{eq:in-out-x} \\
  Y_\outt(t) &= Y_\inn(t) + \sqrt{4\Gqba}\ q(t),\label{eq:in-out-y}
\end{align}
\end{subequations}
The output quadratures of Eqs.~\eqref{eq: input-output} are the quantities that we measure in our experiments and that appear in Eq.~\eqref{eq:7}.
From Eqs.~\eqref{eq:main_eom_q} and \eqref{eq: input-output}, we calculate the following PSD for the homodyne photocurrent
\begin{eqnarray}
  \label{eq:psd_theory}
  S_{ii}^\theta(\Omega) = 1 + S_{\text{imp},\theta}^{-1} |\chi(\Omega)|^2S_{FF}^\text{tot} + 2S_\text{c}^\theta(\Og),
\end{eqnarray}
where we normalized the spectrum to the background noise $S_{\text{imp},\theta}$ (first term on the right-hand side).
The second term arises from the mechanical displacement, with the susceptibility $\chi(\Omega)~=~\Og_m/\left(\Og_m^2-\Og^2-\imu\gm\Omega\right)$ and the total force spectrum $S_{FF}^\text{tot} = 2\left[\Gqba + \gm (\overline{n}+1/2)\right]$.
The sensitivity to the motion is represented by $S_{\text{imp},\theta}^{-1} = 8 \etad \Gqba \sin(\theta)^2$, which is shown in Fig.~\ref{fig:exp}(c).
Finally, the last term in Eq.~\eqref{eq:psd_theory} represents the frequency-dependent correlations between the background and the displacement noise $S_\text{c}^\theta(\Og) = 2\etad\Gqba\text{Re}\left[\chi(\Omega)\right]\sin(2\theta)$.
These correlations are responsible for the asymmetric lineshape in Fig.~\ref{fig:squeezing_1} and for the frequency-dependent squeezing angle in Fig.~\ref{fig:tomography}.

The figure of merit for the degree of ponderomotive squeezing is the measurement efficiency $\eta_\text{meas}=\Gmeas/\Gtot$, where we have introduced for convenience the total decoherence rate $\Gtot = \Gqba + \gm (\overline{n}+1/2)$ and the measurement rate $\Gmeas=\eta_\mathrm{d}\Gqba$.
The correlations lead to significant ponderomotive squeezing when the measurement rate approaches the total decoherence rate, that is $\eta_\mathrm{meas}\sim 1$.
In this limit, the minimum value of the spectrum approaches $S_{ii}^\theta\approx1-\eta_\text{meas}$ close to the resonance frequency \cite{nielsen_multimode_2017}.
We use Eq.~\eqref{eq:psd_theory} to simultaneously fit all the measured spectra, some of which are shown in Fig.~\ref{fig:squeezing_1} (see Supplemental). 
We extract the rates  $\Gtot/(2\pi) = 5.0$~kHz and $\Gmeas/(2\pi)=1.4$~kHz, yielding a measurement efficiency of $\eta_\text{meas}=0.28$. 
These results are consistent with what we previously reported in Ref.~\cite{tebbenjohanns_quantum_2021}. %
These rates allow us to calibrate the displacement measurements in units of zero-point motion.
This calibration technique relies only on the ponderomotive correlations present in the spectra. These spectra, in turn, are calibrated against the optical shot noise, which is easy to quantify experimentally.
The estimated parameters can be also used to compute the theoretical Wigner functions for the modes at $\Omega/(2\pi)=\SI{70.1}{kHz}$ and $\SI{77.1}{kHz}$, whose $X$ and $Y$ marginals are shown in Fig.~\ref{fig:tomography}.
Both the fits of the spectra and the marginals extracted from the theoretical Wigner functions are in good agreement with the measurements.

We have experimentally observed squeezing of light scattered by a levitated nanoparticle and we have fully characterized the optical state with homodyne tomography.
We have measured a reduction of the optical quantum fluctuations by 25\%, which is due to the large measurement efficiency featured by our system.
Notably, we observe ponderomotive squeezing from a single particle in free space, without the need of an optical resonator to enhance the optomechanical coupling.
We model our experiments by using a linearized dipolar treatment of the light-matter interaction.

The ponderomotive correlations present in the scattered field can be readily exploited to provide quantum enhancements in force sensing applications \cite{mason_continuous_2019}, such as gravitational wave detectors based on levitated sensors \cite{arvanitaki_detecting_2013}, in testing fundamental force laws \cite{moore_searching_2021} and in the search for dark matter \cite{carney_mechanical_2021}.

{\it Note added.} We recently became aware of a related independent work by Magrini et al. .

\section*{Acknowledgements}

This research was supported by the Swiss National Science Foundation (SNF) through the NCCR-QSIT programme (grant no. 51NF40-160591), by the European Union’s Horizon 2020 research and innovation programme under grant no. 863132 (iQLev) and by the Q-Xtreme project of the European Research Council under the European Union's Horizon 2020 research and innovation program (grant agreement 951234).
We thank M. L. Mattana for her contributions to the experimental setup.

\bibliographystyle{apsrev4-1}
\bibliography{references_MR} 

\end{document}


\title{Supplemental Material:\\Ponderomotive squeezing of light by a levitated nanoparticle in free space}
\author{Andrei Militaru}
\altaffiliation{\equalcontribution}
\affiliation{\affil}
\author{Massimiliano Rossi}
\altaffiliation{\equalcontribution}
\affiliation{\affil}
\author{Felix Tebbenjohanns}
\altaffiliation{present address: Department of Physics, Humboldt-Universität zu Berlin, 10099 Berlin, Germany}
\affiliation{\affil}
\author{Oriol Romero-Isart}
\affiliation{\affilIQOQI}
\affiliation{\affilITP}
\author{Martin Frimmer}
\affiliation{\affil}
\author{Lukas Novotny}
\affiliation{\affil}
\affiliation{\affilQC}

\maketitle

\tableofcontents
\clearpage
\newpage


\section{Theory}
In this section, we give a complete derivation of the theory used to analyse our experimental data. This theory is based on a quantum optics theoretical framework which assumes a linearized dipolar light-matter interaction. The derivation is based on Refs.~\cite{chang_cavity_2010, romero-isart_optically_2011, rodenburg_quantum_2016, gonzalez-ballestero_theory_2019, magrini_real-time_2021}.

\subsection{System of interest}
\label{sec: system of interest}

We consider a lossless and isotropic dielectric nanoparticle of mass $m$ and real scalar polarizability $\alpha$. The center-of-mass (COM) motion of the nanoparticle is described by its momentum operator  $\bm{p} = (p_x, p_y, p_z)^T$ and by its corresponding position operator $\rr = (x, y, z)^T$. 
In the limit of a particle much smaller than any wavelength of interest, the interaction between the nanoparticle and the surrounding electromagnetic field is given by the dipole interaction $- \bm{d}(\rr) \cdot \bm{E} (\rr)/2$, with $ \bm{d}(\rr) = \alpha \bm{E}(\rr)$ the induced dipole moment and $\bm{E}(\rr)$ the electric field operator evaluated at the nanoparticle's position. The total Hamiltonian can then be described by three terms,
\begin{subequations}
\label{eq: Hamiltonian}
\begin{equation}
H = H_\mathrm{np} + H_\mathrm{f} + H_\mathrm{I},  \label{eq: Htot}
\end{equation}
with the nanoparticle Hamiltonian 
\begin{equation}
H_\mathrm{np} = \frac{\bm{p}^2}{2m}, \label{eq: Hnp}
\end{equation}
the free field Hamiltonian 
\begin{equation}
H_\mathrm{f} = \sum_\varepsilon \intdk \hbar \omega a_\varepsilon^\dagger(\kk) a_\varepsilon(\kk),  \label{eq: Hem}
\end{equation} 
and the interaction Hamiltonian 
\begin{equation}
H_\mathrm{I} =  - \frac{1}{2}\alpha \bm{E}^2(\rr). \label{eq: Hint}
\end{equation}
\end{subequations}
In Eq.~\eqref{eq: Hem}, $\hbar$ is the reduced Planck constant, $\varepsilon$ is the polarization index, $\omega=c\vert\kk\vert$ represents the angular frequency of each free field mode, and $a_\varepsilon(\kk)$ and $a^\dagger_\varepsilon(\kk)$ are the corresponding ladder operators that obey standard bosonic commutation relations: $[a_\varepsilon(\kk), a_{\varepsilon'}(\kk')] = [a^\dagger_\varepsilon(\kk), a^\dagger_{\varepsilon'}(\kk')] = 0$ and $[a_\varepsilon(\kk), a^\dagger_{\varepsilon'}(\kk')] = \delta(\kk-\kk') \delta_{\varepsilon \varepsilon'}$. 
The electric field is evaluated at the nanoparticle's position operator $\rr$, such that Eq.~\eqref{eq: Hint} represents indeed an interaction between $\bm{E}$ and COM motion.
At this point, we deploy the basis of plane waves, such that the annihilation (creation) operators $a_\varepsilon(\kk)$ ($a^\dagger_\varepsilon(\kk)$) in Eq.~\eqref{eq: Hem} refer to a plane wave mode with polarization along the unit vector $\bm{u}^{(\kk)}_\varepsilon$ and wavevector $\kk$. 

In many levitodynamics experiments in free space, the nanoparticle is trapped in the optical potential of a focused laser beam \cite{maurer_quantum_2021, ricci_optically_2017, jain_direct_2016, tebbenjohanns_cold_2019, tebbenjohanns_motional_2020, hebestreit_sensing_2018, hoang_experimental_2018}. Within this framework, we can account for the presence of the trapping beam, referred to in the following as {\it tweezer field}, by splitting the electric field operator $\bm{E}$ into a strong (classical and time dependent) component $\bm{E}_\mathrm{tw}(\rr, t)$ displaced by a coherent amplitude and the quantum fluctuations $\bm{E}_\mathrm{f}(\rr)$
(taken in the Schrödinger picture and hence time independent)
\cite{gonzalez-ballestero_theory_2019}: 
\begin{subequations}
\label{eq: electric field}
\begin{equation}
\label{eq: electric total}
\bm{E}(\rr, t) = \bm{E}_\mathrm{tw}(\rr, t) + \bm{E}_\mathrm{f}(\rr).
\end{equation}
The free field in Eq.~\eqref{eq: electric total} takes the standard form
\begin{equation}
\label{eq: electric free}
\bm{E}_\mathrm{f}(\rr) = \imu \sum_\varepsilon\intdk \vacc\  \bm{u}^{(\kk)}_\varepsilon \left( a_\varepsilon(\kk) \expu^{\imu \kk \cdot \rr} - a^\dagger_\varepsilon(\kk) \expu^{-\imu \kk \cdot \rr}\right),
\end{equation}
with $\varepsilon_0$ being the vacuum permittivity. In this work, the tweezer field is chosen to be an $x$-polarised Gaussian beam whose focus lies at the origin of our frame of reference. In the neighbourhood of the origin, the tweezer field can be written as \cite{novotny_radiation_2017}
\begin{equation}
\label{eq: electric tweezer}
\bm{E}_\mathrm{tw}(\rr, t) =  \frac{E_0\  \exp\left( -\frac{x^2}{w_x^2} - \frac{y^2}{w_y^2} \right)}{\sqrt{1 + \left(\frac{z}{z_\mathrm{R}}\right)^2}}  \ \cos\big(k_0 z-\mathrm{atan}(z/z_\mathrm{R})-\omega_0 t\big)  \bm{u}_x.
\end{equation}
\end{subequations}
In Eq.~\eqref{eq: electric tweezer}, $E_0$ is the field's amplitude at the origin, $w_x$ and $w_y$ are the beam radii along the axes $x$ and $y$, $k_0 = \omega_0/c = 2\pi/\lambda_0$ is the wave number of the beam, $z_\mathrm{R}$ is an effective Rayleigh range, and $\bm{u}_x$ is the unit vector pointing along $x$.

The interaction Hamiltonian in Eq.~\eqref{eq: Hint} contains a term quadratic in the tweezer field $\bm{E}_\mathrm{tw}$, a mixed term and a term quadratic in the fluctuation field $\bm{E}_\mathrm{f}$. We can simplify the interaction hamiltonian by retaining only the terms that depend on the tweezer field, which is assumed to be much stronger than the free field $\bm{E}_\mathrm{f}$.
By neglecting the term quadratic in the free field, we are not taking into account particle-mediated interactions between different electromagnetic modes. This approximation is justified because we consider the limit of an infinitely small nanoparticle \cite{pflanzer_arbitrary_dielectrics}.
%
For finite sized objects, instead, cross terms should be included in Eq.~\eqref{eq: Hem} \cite{pflanzer_arbitrary_dielectrics}, or a more suitable scattering mode basis should be developed (as done in Ref.~\cite{maurer_quantum_2021}). The simplified interaction Hamiltonian is now
\begin{equation}
\label{eq: interaction Hamiltonians}
H_\mathrm{I} \approx - \frac{1}{2}\alpha \bm{E}_\mathrm{tw}\cdot \bm{E}_\mathrm{tw} - \alpha \bm{E}_\mathrm{tw} \cdot\bm{E}_\mathrm{f} =: H_\mathrm{tw} + H_\mathrm{rc}.
\end{equation}
We have defined the first and second terms of Eq.~\eqref{eq: interaction Hamiltonians} respectively as tweezer Hamiltonian $H_\mathrm{tw}$ and recoil Hamiltonian $H_\mathrm{rc}$. The tweezer Hamiltonian is governed by the classical component of the electromagnetic field, while the quantum properties of light are carried by the recoil Hamiltonian. In Secs.~\ref{sec: tweezer Hamiltonian} and~\ref{sec: recoil Hamiltonian} we will show that the tweezer Hamiltonian is responsible for generating the optical trapping potential while the recoil Hamiltonian accounts for the light scattered by the particle and for the related photon recoil. 

\subsection{Optomechanical Hamiltonian}
\label{sec: optomechanical Hamiltonian}

In this section we manipulate the separate terms of Eqs.~\eqref{eq: Hamiltonian} and derive a form of the Hamiltonian that is simple and that can be of direct use for many experiments where the small particle limit is valid. 

\subsubsection{Tweezer Hamiltonian: optical potential}
\label{sec: tweezer Hamiltonian}

The tweezer Hamiltonian is straightforward to analyse. When inserting Eq.~\eqref{eq: electric tweezer} into $H_\mathrm{tw}$ from Eq.~\eqref{eq: interaction Hamiltonians} we get 
\begin{equation}
\label{eq: tweezer Hamiltonian}
H_\mathrm{tw} = -\frac{\frac{1}{2}\alpha E_0^2}{1 + \left(\frac{z}{z_\mathrm{R}}\right)^2} \cos^2\big(k_0 z-\mathrm{atan}(z/z_\mathrm{R})-\omega_0 t\big) \exp\left(- 2 \left(\frac{x^2}{w_x^2} + \frac{y^2}{w_y^2} \right)\right).
\end{equation}
In most experiments, feedback cooling is applied on the COM motion, such that the particle's displacement from the tweezer's focus is very small and we can expand Eq.~\eqref{eq: tweezer Hamiltonian} to second order. Once we expand Eq.~\eqref{eq: tweezer Hamiltonian}, we can neglect the terms oscillating at twice the tweezer frequency, in other words we apply the rotating wave approximation. We are then left with
\begin{subequations}
\label{eq: tweezer Hamiltonian final}
\begin{equation}
\label{eq: optical potential}
H_\mathrm{tw} \approx \frac{1}{2} m \left( \Omega_x^2 x^2 + \Omega_y^2 y^2 + \Omega_z^2 z^2 \right),
\end{equation}
where we defined the three resonance angular frequencies
\begin{equation}
\label{eq: angular resonance frequencies}
\Omega_x^2 = \frac{\alpha E_0^2}{m w_x^2}, \qquad \Omega_y^2 = \frac{\alpha E_0^2}{m w_y^2}, \qquad \Omega_z^2 = \frac{\alpha E_0^2}{2m z_\mathrm{R}^2}.
\end{equation}
\end{subequations}
Equation~\eqref{eq: optical potential} represents the harmonic optical potential generated by the trapping laser. 
Equation~\eqref{eq: optical potential} tells us that the COM motion of the nanoparticle represents a set of three independent harmonic oscillators, one for each axis of motion, whose respective angular frequencies are proportional to the input beam's amplitude $E_0$ and inversely proportional to the beams focal spot size, Eq.~\eqref{eq: angular resonance frequencies}. 

\subsubsection{Recoil Hamiltonian: quantum optical scattering}
\label{sec: recoil Hamiltonian}

The recoil Hamiltonian $H_\mathrm{rc}=-\alpha \bm{E}_\mathrm{tw}\cdot \bm{E}_\mathrm{f}$ in Eq.~\eqref{eq: interaction Hamiltonians} represents the interaction between the trapping beam $\bm{E}_\mathrm{tw}$ and the free field $\bm{E}_\mathrm{f}$, which is mediated by the nanoparticle. This interaction lies at the heart of most levitodynamics experiments: on the one hand, every photon that is rerouted by the particle from the tweezer field to the free field contributes to disturb the particle's motion through photon recoil. On the other hand, as we will see in Sec.~\ref{sec: modes}, measuring the field scattered by the particle allows us to reconstruct its position. The precision of the position measurement is then linked to the photon recoil through the Heisenberg measurement-disturbance relation. 

We consider the common case in which the displacements of the nanoparticle are very small compared to the trapping beam's wavelength. We can then simplify both the tweezer and the free fields in Eqs.~\eqref{eq: electric free} and~\eqref{eq: electric tweezer} by expanding the complex exponentials to first order:
\begin{subequations}
	\label{eq: first order simplifications}
	\begin{align}
	\bm{E}_\mathrm{f}(\rr) &\approx  \sum_\varepsilon\intdk \vacc\  \left[ \imu \left(a_\varepsilon(\kk) - a^\dagger_\varepsilon(\kk) \right)  - \kk\cdot\rr \left(a_\varepsilon(\kk) + a^\dagger_\varepsilon(\kk) \right) \right]\bm{u}^{(\kk)}_\varepsilon, \\
	\bm{E}_\mathrm{tw}(\rr, t) &\approx E_0 \left( \cos\big(\omega_0 t\big) + A k_0 z \sin\big( \omega_0 t \big) \right)\bm{u}_x.
	\end{align}
\end{subequations}
Here, we introduced the geometric factor $A = 1-1/(k_0z_\mathrm{R})$ to account for the Gouy phase in the neighbourhood of the origin. We then substitute Eqs.~\eqref{eq: first order simplifications} into the recoil Hamiltonian $H_\mathrm{rc}$ in Eq.~\eqref{eq: interaction Hamiltonians} and retain only terms linear in the position. The result reads
	\begin{align}
	\label{eq: Hrc}
	H_\mathrm{rc} &\approx - \imu \cos\big(\omega_0 t \big) \sum_\varepsilon\intdk G(\kk, \varepsilon) \left(a_\varepsilon(\kk) - a^\dagger_\varepsilon(\kk)\right) \nonumber \\ 
	&\qquad + \cos \big(\omega_0 t \big) \sum_\varepsilon\intdk G(\kk, \varepsilon)\  \left( a_\varepsilon(\kk) + a^\dagger_\varepsilon(\kk) \right) \kk\cdot\rr \nonumber \\
	&\qquad -\imu \sin \big(\omega_0 t \big) \sum_\varepsilon\intdk G(\kk, \varepsilon)\  \left( a_\varepsilon(\kk) - a^\dagger_\varepsilon(\kk) \right) Ak_0 z.
	\end{align}
In Eq.~\eqref{eq: Hrc}, the function $G(\kk, \varepsilon)$ is defined in analogy to Ref.~\cite{gonzalez-ballestero_theory_2019} as
\begin{equation}
\label{eq: coupling G}
G(\kk, \varepsilon) = \alpha E_0 \vacc (\bm{u}^{(\kk)}_\varepsilon \cdot \bm{u}_x),
\end{equation}
and expresses the strength of the coupling between the tweezer and each plane wave mode as a function of the wavevector $\kk$ and of the polarization $\varepsilon$.
The time dependence of Eq.~\eqref{eq: Hrc} can be eliminated by moving to a frame rotating at the tweezer frequency, i.e., we substitute $a_\varepsilon(\kk) \to a_\varepsilon(\kk) \exp\left( -\imu \omega_0 t \right)$ and $a^\dagger_\varepsilon(\kk) \to a^\dagger_\varepsilon(\kk) \exp\left(  \imu \omega_0 t \right)$, and neglecting the fast rotating terms:
	\begin{align}
\label{eq: Hrc RWA}
H_\mathrm{rc} &\approx - \imu  \sum_\varepsilon\intdk \frac{G(\kk, \varepsilon)}{2} \left(a_\varepsilon(\kk) - a^\dagger_\varepsilon(\kk)\right) \nonumber \\ 
&\quad - \sum_\varepsilon\intdk \frac{G(\kk, \varepsilon)}{2}\  \left( a_\varepsilon(\kk) + a^\dagger_\varepsilon(\kk) \right) \left( \kk\cdot\rr - Ak_0 z \right).
\end{align}
Equation~\eqref{eq: Hrc RWA} is the main result of this subsection. It consists of two terms. The first one, position independent, displaces the free field modes to a coherent state whose strength as a function of orientation is proportional the coupling function $G(\kk, \varepsilon)$. This first term represents the quantum description of the Rayleigh scattering of the nanoparticle. The second term describes instead a position dependent displacement of the free field, whose radiation pattern is different from the Rayleigh scattering due to the presence of the term $\kk$. In the next section, we explicitly solve the summation over all modes and show that only three free space modes are sufficient to describe the optomechanical interaction present in Eq.~\eqref{eq: Hrc RWA}.

\subsubsection{Interacting modes}
\label{sec: modes}

So far in this work, we have used the plane-wave basis for the free space modes. Within our small-particle approximation, this plane-wave basis provides a convenient mathematical tool that has allowed us to derive the simplified recoil Hamiltonian of Eq.~\eqref{eq: Hrc RWA}. In Eq.~\eqref{eq: Hrc RWA}, every COM degree of freedom is coupled to every mode $a_\varepsilon(\kk)$ with a coupling strength that depends on the orientation of the plane wave and on its frequency. In this subsection, we
effectively solve the integrals over plane wave orientation in Eq.~\eqref{eq: Hrc RWA} and
introduce a new mode basis, consisting of what we define as {\it interacting modes}. We use the interacting modes
to derive the main theoretical result of this work. 

When considering Eq.~\eqref{eq: Hrc RWA}, we can see that both terms involve a summation over all plane-wave modes with different weights. For the $x$, $y$ and $z$ dependent terms, we find weights given by $G(\kk, \varepsilon) k_x$, $G(\kk, \varepsilon) k_y$, and $G(\kk, \varepsilon) (k_z - k_0A)$ respectively. It is then appealing to define 
interacting modes as
\begin{equation}
\label{eq: interacting modes propto}
a_j(\omega) \propto \sum_{\varepsilon}\intdk G(\kk, \varepsilon) \delta\left(k - \frac{\omega}{c}\right) \tilde{k}_j a_\varepsilon(\kk),
\end{equation}
where the Dirac $\delta$ selects the modes with angular frequency $\omega$ and $\tilde{k}_j = k_j - A\delta_{zj}$. By choosing suitable numerical prefactors to guarantee
\begin{equation}
\label{eq: commutator interacting modes}
\left[a_j(\omega), a^\dagger_{j'}(\omega')\right] = \delta_{jj'}\delta(\omega-\omega'),
\end{equation}
and using Eq.~\eqref{eq: coupling G} we reach the expressions
\begin{subequations}
	\label{eq: xyz}
\begin{align}
a_x(\omega) &= -\sqrt{\frac{15c}{8\pi\omega^2}} \sum_\varepsilon \intdk \delta\left(k - \frac{\omega}{c}\right) \frac{k_x}{k}  \left(\bm{u}_x \cdot \bm{u}^{(\kk)}_\varepsilon\right) a_\varepsilon(\kk), \label{eq: optomechanical x}\\
a_y(\omega) &= -\sqrt{\frac{15c}{16\pi\omega^2}} \sum_\varepsilon \intdk \delta\left(k - \frac{\omega}{c}\right) \frac{k_y}{k}  \left(\bm{u}_x \cdot \bm{u}^{(\kk)}_\varepsilon\right) a_\varepsilon(\kk), \label{eq: optomechanical y}\\
a_z(\omega) &=  -\sqrt{\frac{15c}{8\pi\omega^2(2+5A^2)}}\sum_\varepsilon \intdk \delta\left(k - \frac{\omega}{c}\right)  \left( \frac{k_z - k_0A}{k} \right) \left(\bm{u}_x \cdot \bm{u}^{(\kk)}_\varepsilon\right) a_\varepsilon(\kk), \label{eq: optomechanical z}
\end{align}
\end{subequations}
where the negative signs have been chosen for later notational convenience. 
%
With Eqs.~\eqref{eq: xyz}, the commutator condition in Eq.~\eqref{eq: commutator interacting modes} is indeed fulfilled. We show this explicitly for the $x$ interacting mode (the verification of the other modes proceeds analogously). Using $[a_\varepsilon(\kk), a_{\varepsilon'}(\kk')] = \delta_{\varepsilon\varepsilon'}\delta(\kk - \kk')$, we get
\begin{align}
    \label{eq: explicit commutator}
    [a_x(\omega), a_x^\dagger(\omega')] &= \frac{15c}{8\pi\omega^2} \intdk \delta\left( k - \frac{\omega}{c}\right) \delta\left( k - \frac{\omega'}{c}\right) \frac{k_x^2}{k^2} \sum_\varepsilon \left( \bm{u}_x\cdot\bm{u}_\varepsilon^{(\kk)}\right) \nonumber \\ 
    &= \delta(\omega-\omega') \frac{15c^2}{8\pi\omega^2} \int_{4\pi}\dd\Omega\,k_x^2 \sum_\varepsilon \left( \bm{u}_x\cdot\bm{u}_\varepsilon^{(\kk)}\right),
\end{align}
where $\mathrm{d}\Omega$ represents the infinitesimal solid angle.
We can solve the sum over the polarization degrees of freedom using the closure relation \cite{gonzalez-ballestero_theory_2019}
\begin{equation}
    \label{eq: polarization closure relation}
    \sum_\varepsilon \left( \bm{u}^{(\kk)}_\varepsilon \cdot \bm{v}\right)^2 = \bm{v} \cdot \bm{v} - \frac{ (\bm{v} \cdot \kk)^2}{k^2},
\end{equation}
with the help of which we write Eq.~\eqref{eq: explicit commutator} as
\begin{align}
    [a_x(\omega), a_x^\dagger(\omega')] &= \delta(\omega - \omega')\frac{15}{8\pi} \int_{4\pi}\dd\Omega\,\frac{k_x^2}{k^2}\left( 1 - \frac{k_x^2}{k^2} \right) \nonumber \\
    &= \delta(\omega - \omega')\frac{15}{8\pi} \int_0^{2\pi}\dd\varphi\int_0^\pi\dd\theta\,\sin^3(\theta)\cos^2(\varphi)\left( 1 - \cos^2(\varphi)\sin^2(\theta)\right) \nonumber \\
    &= \delta(\omega - \omega'),
\end{align}
where we made use of the polar and azimuthal spherical coordinates, respectively $\theta$ and $\varphi$.

In addition to the interacting modes of Eqs.~\eqref{eq: xyz}, the first line of Eq.~\eqref{eq: Hrc RWA} excites the $x$-oriented dipole mode
\begin{equation}
    \label{eq: dipole mode}
    a_0(\omega) =  \sqrt{\frac{3c}{8\pi\omega^2}}\sum_\varepsilon \intdk \delta\left(k - \frac{\omega}{c}\right)  \left(\bm{u}_x \cdot \bm{u}^{(\kk)}_\varepsilon\right) a_\varepsilon(\kk),
\end{equation}
which describes Rayleigh scattering, as we will show in Sec.~\ref{sec: information patterns}. While the dipole mode in Eq.~\eqref{eq: dipole mode} does not affect the COM degrees of freedom directly [the first line of Eq.~\eqref{eq: Hrc RWA} contains no position-dependent factor], it does impact the dynamics of the $z$ motion of the nanoparticle. Its effect can be understood by noticing that \footnote{We define $a^\dagger_j(\omega)=\left[a_j(\omega)\right]^\dagger$}
\begin{equation}
    \label{eq: Rayleigh overlap}
    \left[a_0(\omega), a^\dagger_z(\omega')\right] = \sqrt{\frac{5A^2}{2+5A^2}}\  \delta(\omega-\omega'),
\end{equation}
in other words the $z$-interacting mode and the Rayleigh scattering mode are not orthogonal. In order to limit our analysis to the COM motion and to the interacting modes only, we can decompose the Rayleigh scattering mode in two components,
\begin{equation}
    \label{eq: Rayleigh decomposition}
    a_0(\omega) = \sqrt{\frac{5A^2}{2+5A^2}}\ a_z(\omega) + g_\mathrm{n} a_\mathrm{n}(\omega),
\end{equation}
where the first one accounts for Eq.~\eqref{eq: Rayleigh overlap} and the second one---whose exact expression is irrelevant for our purpose---is orthogonal to all interacting modes and can be neglected because it does not interact with the COM motion.

Using the interacting modes in Eqs.~\eqref{eq: xyz} and the coupling functions
\begin{subequations}
	\label{eq: g coupling constants}
	\begin{align}
	g_x(\omega) &= \frac{\alpha E_0 k}{\hbar} \sqrt{\frac{\pi \hbar \omega^3}{15\varepsilon_0 (2\pi c)^3}}, \\
	g_y(\omega) &= \frac{\alpha E_0 k}{\hbar} \sqrt{\frac{2\pi \hbar \omega^3}{15\varepsilon_0 (2\pi c)^3}}, \\
	g_z(\omega) &= \frac{\alpha E_0 k}{\hbar} \sqrt{\frac{(2+5A^2)\pi \hbar \omega^3}{15\varepsilon_0 (2\pi c)^3}}, \\
	g_\mathrm{rp}(\omega) &= \frac{\alpha E_0 k}{\hbar} \sqrt{\frac{\pi \hbar \omega^3}{3\varepsilon_0 (2\pi c)^3}} \sqrt{\frac{5A^2}{2 + 5A^2}}.
	\end{align}
\end{subequations}
we can write then the recoil Hamiltonian simply as
\begin{align}
	\label{eq: optomechanical Hamiltonian}
	H_\mathrm{rc} =& \sum_{j \in \{x, y, z\}} \hbar  \int_\mathbb{R}\dd\omega\ \left[-g_j(\omega) r_j \left( a_j(\omega) + a_j^\dagger(\omega) \right) 
	+ \imu \delta_{zj} \frac{g_\mathrm{rp}}{k}(\omega) \left( a_z(\omega) - a_z^\dagger(\omega) \right)\right],
\end{align}
where $r_j=\rr \cdot \bm{u}_j$ represents an alternative notation for the position operator along the $j$ axis.
Note that we extended the lower bound of the integrals over the angular frequency $\omega$ to $-\infty$ rather then $-\omega_0$. This approximation is convenient and valid as long as the particle's dynamics is much slower than the tweezer frequency $\omega_0$, in other words as long as the Markovian approximation is valid \cite{gardiner_input_1985}.

The resulting recoil Hamiltonian Eq.~\eqref{eq: optomechanical Hamiltonian} is known in literature as {\it optomechanical interaction} \cite{clerk_introduction_2010, aspelmeyer_cavity_2014-1}. Every position degree of freedom $r_j$ interacts with infinitely many modes  $a_j(\omega)$ (one for every angular frequency $\omega$) belonging to only one spatial mode, which is the reason why we named these spatial modes {\it interacting modes}. The interaction is such that each interacting modes is coherently displaced by an amount proportional to the nanoparticle's COM displacement, and the proportionality factors are detailed in Eqs.~\eqref{eq: g coupling constants}. Moreover, the $z$ interacting mode presents a position-independent populating term [second part of Eq.~\eqref{eq: optomechanical Hamiltonian}], which in turn affects $z$ through their optomechanical interaction. This position-independent term is responsible for the radiation pressure that the tweezer beam exerts on the nanoparticle, as we will explain in Sec.~\ref{sec: input output}.

\subsubsection{Information patterns}
\label{sec: information patterns}

We have seen that the interaction between the mechanical degrees of the nanoparticle and the free electromagnetic field becomes considerably simpler when described in terms of the interacting modes. In this section, we further investigate their meaning and physical form.

To this end, let us conduct a thought experiment. We use the spherical coordinates $\theta$ and $\varphi$, given by the polar angle with respect to the $z$ axis and the azimuthal angle, respectively. We can then define the photon number operator $N$ as 
\begin{equation}
    \label{eq: photon number}
    N = \sum_\varepsilon\intdk  a^\dagger_\varepsilon(\kk)a_\varepsilon(\kk) = \int_{4\pi}\dd\Omega\  \underbrace{\sum_\varepsilon  \int_{\mathbb{R}^+} \dd k\ k^2 a^\dagger_\varepsilon(\kk)a_\varepsilon(\kk)}_{\mathcal{N}(\theta, \varphi)},
\end{equation}
where $\Omega$ refers to the solid angle along $(\theta, \varphi)$. In Eq.~\eqref{eq: photon number}, $N$ quantifies the number of photons present in a state independent from their frequency, orientation or polarization. As shown in Eq.~\eqref{eq: photon number}, it is possible to define the density of photons $\mathcal{N}$ {\it along a direction} $(\theta, \varphi)$. Integrating $\mathcal{N}$ over the full solid angle yields then the total number of photons \footnote{The integral over the total solid angle takes the standard definition $\int_{4\pi}\dd\Omega = \int_0^\pi\dd\theta\ \int_0^{2\pi}\dd\varphi$}. Equipped with the concept of density of photons $\mathcal{N}$, we can introduce the radiation pattern  $\rho$ corresponding to a state $\vert \psi \rangle$ of the optical field as
\begin{equation}
    \label{eq: radiation pattern}
    \rho(\theta, \varphi) := \langle \mathcal{N}(\theta, \varphi) \rangle = \langle \psi \vert \mathcal{N}(\theta, \varphi) \vert \psi \rangle.
\end{equation}
Equation~\eqref{eq: radiation pattern} is our key to understand the interacting modes. We consider the optical vacuum state $\vert 0 \rangle$ and excite one of the interacting modes with a coherent state. Rather that focusing on a specific frequency of the interacting modes, $a_j(\omega)$, it is beneficial to consider the multimode operator
\begin{equation}
    \label{eq: multimode a}
    a_j^\dagger = \int_{\mathbb{R}^+} \dd\omega\ f(\omega) a_j^\dagger(\omega),
\end{equation}
whose commutator with $a_\varepsilon(\kk)$ is
\begin{equation}
    \label{eq: commutator lambda}
    [a_\varepsilon(\kk), a^\dagger_j] = \frac{c}{N_j} \frac{\tilde{k}_j}{k} f(kc) \left( \bm{u}_x\cdot\bm{u}_\varepsilon^{(\kk)}\right) =: \Lambda_j(\varepsilon, \kk),
\end{equation}
where 
\begin{equation}
\label{eq: prefactors}
N_j^2 = \frac{8\pi\omega^2}{15c} \begin{pmatrix}
1 \\ 2 \\ 2+5A^2
\end{pmatrix},
\end{equation}
 and $\tilde{k}_j = k_j - \delta_{zj}A$. The weight function $f$, normalized such that $\int_{\mathbb{R}^+}\dd\omega\ \vert f \vert^2 =1$, can be chosen arbitrarily narrow or wide around a frequency of interest. Using the multimode operator in Eq.~\eqref{eq: multimode a}, rather than Eqs.~\eqref{eq: xyz}, allows us to bypass the singular nature of the commutator of continuous mode operators. We can now excite the interacting mode with a coherent state of amplitude $\beta$. Our excited interacting mode takes then the form
\begin{equation}
\label{eq: excited state}
\vert \beta \rangle_j = \mathcal{D}_j(\beta) \vert 0 \rangle = \expu^{\beta a_j^\dagger - \beta^*a_j} \vert 0 \rangle = \expu^{-\vert\beta\vert^2/2} \expu^{\beta a^\dagger_j} \vert 0 \rangle.
\end{equation}
We can see from Eq.~\eqref{eq: commutator lambda} and Eq.~\eqref{eq: excited state} that \footnote{Use $[a_\varepsilon, a_j^{\dagger n}] = \Lambda_j(\varepsilon, \kk) n a_j^{\dagger (n-1)}$ and the Taylor expansion of $\exp(\beta a_j^\dagger)$.}
\begin{equation}
    \label{eq: coherent state property}
    a_\varepsilon(\kk) \vert \beta \rangle_j = \beta \Lambda_j(\varepsilon, \kk) \vert \beta \rangle_j,
\end{equation}
which we substitute in Eq.~\eqref{eq: radiation pattern} to obtain 
\begin{subequations}
	\label{eq: differential power}
	\begin{align}
	\rho_x &=  \vert\beta\vert^2 \frac{15}{8\pi}\sum_\varepsilon  \left(\bm{u}_x \cdot \bm{u}^{(\kk)}_\varepsilon\right)^2 \frac{k_x^2}{k ^2}, \\
	\rho_y &=  \vert\beta\vert^2 \frac{15}{16\pi}\sum_\varepsilon  \left(\bm{u}_x \cdot \bm{u}^{(\kk)}_\varepsilon\right)^2 \frac{k_y^2}{k ^2}, \\
	\rho_z &= \vert\beta\vert^2 \frac{15}{8\pi (2 + 5A^2)} \sum_\varepsilon \left(\bm{u}_x \cdot \bm{u}^{(\kk)}_\varepsilon\right)^2\left( \frac{k_z}{k} - A \right)^2,
	\end{align}
\end{subequations}
with the angle dependence implicitly contained in $\kk$ and $\bm{u}_\varepsilon^{(\kk)}$. 
Using the polarization closure relation Eq.~\eqref{eq: polarization closure relation} we get
\begin{subequations}
	\label{eq: information patterns}
	\begin{align}
	\rho_x &= \vert\beta\vert^2\frac{15}{8\pi} \big(1 - \cos^2\varphi \sin^2 \theta \big) \cos^2\varphi \sin^2 \theta, \\
	\rho_y &= \vert\beta\vert^2\frac{15}{16\pi} \big(1 - \cos^2\varphi \sin^2 \theta \big) \sin^2\varphi \sin^2 \theta, \\
	\rho_z &= \vert\beta\vert^2\frac{15}{8\pi (2 + 5A^2)} \big(1 - \cos^2\varphi \sin^2 \theta \big) \left( \cos \theta - A \right)^2.
	\end{align}
\end{subequations}
When looking at Eqs.~\eqref{eq: information patterns} we see that the radiation patterns of the interacting modes are not new to the community. In fact, these patterns have been already derived in a semiclassical calculation \cite{tebbenjohanns_optimal_2019}. In the context of Ref.~\cite{tebbenjohanns_optimal_2019}, $\rho_j(\theta, \varphi)$ represented the information density about the position $r_j$ of a dipolar scatterer contained along the direction $(\theta, \varphi)$. In this work, these information patterns play the role of orthonormal electromagnetic modes that interact respectively with the position $r_j$ only. Within the current framework, it is natural that an optimal measurement of $r_j$ must involve the radiation pattern $\rho_j$. 

A second way to understand Eqs.~\eqref{eq: information patterns}, which is mathematically simpler but further away from a classical intuition, is to populate the interacting modes with one photon, in other words to use $\vert \psi \rangle = a_j^\dagger \vert 0 \rangle$ in Eq.~\eqref{eq: radiation pattern}. The result coincides once again with Eqs.~\eqref{eq: information patterns} for $\beta = 1$. The information patterns can be then interpreted as the probability to find a photon along $(\theta, \varphi)$ when the interacting mode contains one photon.

Finally, we can repeat the same procedure on the dipole mode, Eq.~\eqref{eq: dipole mode}, and derive its radiation pattern
\begin{equation}
\label{eq: dipole radiation pattern}
\rho_0 = \vert\beta\vert^2\frac{3}{8\pi} \big(1 - \cos^2\varphi \sin^2 \theta \big),
\end{equation}
which coincides indeed with the dipole radiation from classical electrodynamics \cite{novotny_radiation_2017}.

\subsubsection{Total Hamiltonian}
\label{sec: final Hamiltonian}

We conclude this section by putting together the results achieved so far. In Sec.~\ref{sec: tweezer Hamiltonian} we have seen how the tweezer Hamiltonian generates a harmonic potential for the nanoparticle. In addition to this, the nanoparticle mediates a coupling between the positions $r_j$ and the free space electromagnetic mode. Importantly, each position $r_j$ interacts with only one spatial electromagnetic mode when the basis of interacting modes is chosen \footnote{Note that we use the expression {\it interacting mode} to refer to the spatial shape of the optical modes described by Eqs.~\eqref{eq: xyz}. Each interacting mode consists then of infinitely many frequency modes.}. The total optomechanical Hamiltonian that describes the problem consists thus of three mutually commuting terms:
\begin{subequations}
\begin{equation}
\label{eq: total Hamiltonian}
H = H_x + H_y + H_z,
\end{equation}
where each of the Hamiltonians $H_j$ takes the form
\begin{align}
\label{eq: Hj continuous}
H_j &= \frac{p_j^2}{2m} + \frac{1}{2} m \Omega_j^2 r_j^2 + \int_\mathbb{R}\dd\omega\ \hbar \omega a_j^\dagger(\omega) a_j(\omega)  - \hbar g_j r_j \int_\mathbb{R}\dd \omega\  \left(a_j(\omega)  + a^{\dagger}_j(\omega) \right) \nonumber \\
&\qquad+ \imu \delta_{jz} \hbar \frac{g_\mathrm{rp}}{k} \int_\mathbb{R}\dd\omega\ \left(a_j(\omega)  - a^{\dagger}_j(\omega) \right),
\end{align}
\end{subequations}
and where the Markovian approximation has been applied, i.e. the coupling constants are evaluated at the tweezer frequency: $g_j(\omega) \approx g_j(\omega_0) = g_j$(and analogously for $g_\mathrm{rp}$). This approximation is valid as long as the dynamics of the particle is much slower than the tweezer frequency. 

Equation~\eqref{eq: Hj continuous} contains many terms, nevertheless they can be readily understood. The first three terms contain the harmonic dynamics of the nanoparticle and the free evolution of the $j$-interacting mode. This $j$-interacting mode interacts with the position $r_j$ through a bilinear Hamiltonian given by the fourth term. Finally, the last line of Eq.~\eqref{eq: Hj continuous} is nonzero only for $H_z$ and represents a coherent population of the $z$-interacting mode. Because of the fourth term of Eq.~\eqref{eq: Hj continuous}, the coherent drive translates into a constant force pushing the particle along the $z$ axis, in other words this term describes the radiation pressure (see Sec.~\ref{sec: input output}). Within this framework, the radiation pressure can be understood as a consequence of the fact that the Rayleigh scattering mode has a finite mode overlap with the $z$-interacting mode.

Equation~\eqref{eq: Hj continuous} is cast in a form that is ideally suited to derive the quantum Langevin description of the system, which we address in the following section.

\subsection{Quantum Langevin equations}
\label{sec: input output}
In this section, we focus the attention to the dynamics related to the motion along the $z$ axis only. The equations of motion for the transverse modes $x$ and $y$ can be derived in an analogous fashion.
From $H_z$ in Eq.~\eqref{eq: Hj continuous}, we obtain the following Heisenberg equations of motion 
\begin{subequations}
\begin{align}
  \dot{z} &= p_z/m, \label{eq: heisenberg z}\\
  \dot{p}_z &= - m\Omega_z^2 z + \hbar g \int_\mathbb{R}\dd\omega\ \left( a(\omega) + a^\dagger(\omega) \right), \label{eq: heisenberg p}\\
  \dot{a}(\omega, t) &= -\imu \omega a(\omega,t) + \tilde{g}(t), \label{eq:2}
  \end{align}
\end{subequations}
where $\tilde{g}(t)=g_\text{rp}/k + \imu g z(t)$ is the drive of the field and where we have neglected the $z$ subscripts for conciseness.
Assuming the optical mode to have an initial condition $a(\omega, t_-)$ in the past ($t_-<t$), the solution of Eq.~\eqref{eq:2} is \cite{gardiner_input_1985}
\begin{eqnarray}
  \label{eq:inputsol}
  a(\omega, t) = a(\omega, t_-) \expu^{\imu \omega(t_- - t)} + \int_{t_-}^t \dd s\ \tilde{g}(s) \expu^{\imu \omega(s-t)}.
\end{eqnarray}
%
Equivalently, by choosing the boundary conditions $a(\omega, t_+)$ in the future ($t_+>t$) we have
\begin{eqnarray}
  \label{eq:outputsol}
  a(\omega,t) = a(\omega, t_+) \expu^{\imu \omega(t_+-t)} -  \int_{t}^{t_+} \dd s\ \tilde{g}(s) \expu^{\imu \omega(s-t)}.
\end{eqnarray}
%
It is useful at this point to define the input (output) field as
\begin{subequations}
  \label{eq:4}
  \begin{align}
  a_\inn(t) &= \frac{1}{\sqrt{2\pi}}\int_{-\infty}^\infty \dd\omega\  a(\omega,t_-) \expu^{\imu\omega(t_- -t)}, \\
  a_\outt(t) &= \frac{1}{\sqrt{2\pi}}\int_{-\infty}^\infty \dd\omega\  a(\omega,t_+) \expu^{\imu\omega(t_+ -t)},
  \end{align}
\end{subequations}
which has the commutation relations $[a_i(t), a^\dagger_j(t')]=\delta_{ij}\delta(t-t')$, where $j=\{\text{in, out}\}$. The input (output) field represents the free evolution of the optical field before (after) the interaction with the mechanical degrees of freedom occurred and propagated until the present time $t$ assuming no interaction. For later convenience, we describe the field in terms of Hermitian quadrature amplitudes (later simply termed quadratures) according to $\aout = \left(X_\text{out}+\imu Y_\text{out}\right)/\sqrt{2}$.

By integrating both hand sides of  Eqs.~\eqref{eq:inputsol} and~\eqref{eq:outputsol} and using Eqs.~\eqref{eq:4} we can obtain the following input-output relations for the quadratures \cite{gardiner_input_1985}:
\begin{subequations}\label{eq:5}
  \begin{align}
    X_\text{out}(t) &= X_\text{in}(t) + \sqrt{2\Gamma_\mathrm{R}},\label{eq:io_x}\\
    Y_\text{out}(t) &= Y_\text{in}(t) + \sqrt{4\Gqba} q(t),\label{eq:io_y}
  \end{align}
\end{subequations}
where we have introduced the scattering rate $\Gamma_\mathrm{R}= 2\pi g^2_\text{rp}/k^2$ and the quantum backaction decoherence rate $\Gqba=2\pi g^2_z \zzpf^2$, with the zero point fluctuations $\zzpf^2=\hbar/(2m\Oz)$ and the dimensionless position $q=z/(\sqrt{2}\zzpf)$.

By substituting Eq.~\eqref{eq:inputsol} in Eqs.~\eqref{eq: heisenberg z} and \eqref{eq: heisenberg p}, the equations of motion for the mechanical position and momentum read
\begin{subequations}\label{eq:eom_mech}
  \begin{align}
  \dot{q} &= \Oz p,\label{eq:eom_q}\\
  \dot{p} &= -\Oz q +\sqrt{2\Gqba\Gamma_\mathrm{R}} + \sqrt{4\Gqba} X_\text{in}(t),\label{eq:eom_p}
  \end{align}
\end{subequations}
where $p = p_z/(\sqrt{2}\pzpf)$, with $\pzpf^2 = m\Omega_z \hbar/2$.
The tweezer field exerts a constant radiation pressure force on the particle, which is captured by the second term in Eq.~\eqref{eq:eom_p}.
With the use of Eqs.~\eqref{eq: g coupling constants}, we can appreciate that this constant force can be rewritten as $PA/c$, where $P = (\omega_0^4 \alpha^2 E_0^2)/(12\pi \varepsilon_0 c^3)$ is the optical power radiated by the induced dipole. 
In the following, we account for the radiation pressure by shifting the equilibrium position of the nanoparticle to $q_\text{eq}=\sqrt{2\Gqba\Gamma_\text{rp}}/\Oz$. 
The quantum fluctuations associated with the input optical field are depicted by the last term in Eq.~\eqref{eq:eom_p}, and act as a random force noise .
We notice that there is no dissipation term associated with these fluctuations.
This is in contrast with a relativistic treatment, which predicts the presence of damping via radiative dissipation~\cite{novotny_radiation_2017}.

Note that the Hamiltonian used in the main text was taken in the interaction picture with respect to the free optical field. In this picture, with $t_-$ chosen as the initial time and neglecting the static radiation pressure, the Hamiltonian $H^{(\mathrm{I})}$ takes the form
\begin{equation}
    \label{eq: interaction picture}
    H^{(\mathrm{I})} = \frac{1}{2}\hbar\Omega_z\left(p^2 + q^2\right) - \sqrt{4\Gamma_\mathrm{qba}}\ q X_\inn(t),
\end{equation}
whose last term is the interaction Hamiltonian reported in the main text.
Equation~\eqref{eq: interaction picture} is time dependent, however it commutes with itself at different times. With Eq.~\eqref{eq: interaction picture}, Eqs~\eqref{eq:eom_mech} follow directly from the Heisenberg equations of motion for $q$ and $p$.
%
\subsubsection{Heisenberg measurement-disturbance relation}
Equations~\eqref{eq:5} and \eqref{eq:eom_mech} describe a continuous displacement quantum measurements \cite{jacobs_straightforward_2006, wiseman_quantum_2010}.
Similar equations can be derived for the transverse motion, with $\Gqba^{(j)} = 2\pi g_j^2 j^2_\text{zpf}$ and $j={x,y}$.
The interacting modes along these direction are symmetric (differently from the $z$ mode), thus do not feature any constant radiation pressure force.
Because the input optical field is in the vacuum state, the quadratures have the following non-zero correlations
\begin{eqnarray}\label{eq:input_quad_corr}
\langle X_\inn(t) X_\inn(t') \rangle = \langle Y_\inn(t) Y_\inn(t') \rangle &= \frac{1}{2} \delta(t-t').
\end{eqnarray}
This has two profound implications for the measurement.

Firstly, the estimation of the particle position from a phase quadrature measurement of the output field, Eq.~\eqref{eq:io_y}, is affected by measurement imprecision due to the fluctuating input phase quadrature.
We calibrate this measurement imprecision in displacement units by dividing by $\sqrt{2\Gqba/j_\text{zpf}}$.
From Eqs.~\eqref{eq:io_y} and \eqref{eq:input_quad_corr}, we find the following imprecision power spectral densities (PSDs) \footnote{Given a hermitian operator $\mathcal{O}(t)$, we define its spectral density as $S_{\mathcal{O}\mathcal{O}}(\Omega) = 1/(2\pi) \int_\mathbb{R}\dd\tau\ e^{i\Omega \tau}\langle \overline{ \mathcal{O}(t) \mathcal{O}(t+\tau)}\rangle$, where the bar indicates symmetrization.}
\begin{equation}
\label{eq: all uncertainty measurements}
\begin{pmatrix} \overline{S}^{(x)}_\mathrm{imp} \\ \overline{S}^{(y)}_\mathrm{imp} \\ \overline{S}^{(z)}_\mathrm{imp} \end{pmatrix} = \frac{5}{8\pi} \frac{1}{k_0^2} \frac{\hbar \omega_0}{ P} \begin{pmatrix} 1 \\ 1/2 \\ \left( 2/5 + A^2 \right)^{-1} \end{pmatrix},
\end{equation}
where we made again use of the power scattered by the dipole $P = (\omega_0^4 \alpha^2 E_0^2)/(12\pi \varepsilon_0 c^3)$. These spectra are inversely proportional to the number of scattered photons, $P/(\hbar \omega_0)$, and to the (square of the) wave number $k_0$. 
This makes sense, since the uncertainty is reduced if the number of probing photons is increased and since the resolution of each photon increases with lower wavelengths.

Secondly, the vacuum fluctuations in the amplitude quadrature randomly drive the mechanical motion, as shown in Eq.~\eqref{eq:eom_p}.
This fluctuating force, known as photon recoil \cite{jain_direct_2016, gonzalez-ballestero_theory_2019, tebbenjohanns_optimal_2019}, is a form of quantum backaction, a ubiquitous aspect of quantum measurements. 
The photon recoil PSDs are
\begin{equation}
\label{eq: all photon recoils}
\begin{pmatrix} \overline{S}^{(x)}_{FF} \\ \overline{S}^{(y)}_{FF} \\ \overline{S}^{(z)}_{FF}  \end{pmatrix} = \frac{1}{5}\frac{\hbar^2 k_0^2}{2\pi} \frac{P}{\hbar \omega_0} \begin{pmatrix} 1 \\ 2 \\ 2+5A^2 \end{pmatrix}.
\end{equation}
The strength of this backaction increases with the number of the scattered photons and the momentum carried by each of them.

The imprecision and quantum backaction, respectively Eqs.~\eqref{eq: all uncertainty measurements} and \eqref{eq: all photon recoils}, are not independent.
In fact, their product fulfills the relation
\begin{equation}
\label{eq: heisenberg uncertainty}
\overline{S}^{(j)}_\mathrm{imp}\,\overline{S}^{(j)}_{FF} = \frac{1}{4\pi^2} \frac{\hbar^2}{4}.
\end{equation}
This is an example of the Heisenberg measurement-disturbance relation \cite{braginsky_quantum_1992}, and stems from the quantum nature of the measuring system, that is the electromagnetic field.

\subsubsection{Coupling to a thermal bath}

In practice, the mechanical system is also driven by fluctuating forces other than the photon recoil.
We can take these additional fluctuating forces into account by including an effective thermal bath.
In the absence of any other forces, the mechanical motion thermalizes to the same average phonon occupancy of this bath, $\overline{n}_\mathrm{th}$, at a rate $\gm$.
The equations of motion become then
\begin{subequations}\label{eq:eom_mech_2}
  \begin{align}
  \dot{q} &= \Oz p,\label{eq:eom_q_2}\\
  \dot{p} &= -\Oz q - \gm p +\xi(t) + \sqrt{4\Gqba} X_\text{in}(t),\label{eq:eom_p_2}
  \end{align}
\end{subequations}
where we have introduced an effective thermal force with correlations $\langle\overline{\xi(t)\xi(t')}\rangle=2\gm\left(\overline{n}_\mathrm{th}+1/2\right)\delta(t-t')$ \cite{giovannetti_phase-noise_2001}.

Equations~\eqref{eq:5} and~\eqref{eq:eom_mech_2} correspond to Eqs.~(3) and~(4) of the main text.

\section{Experimental details}
In this section, we provide more details about experimental methods and calibration, the spectral fittings and the reconstruction of the Wigner functions.

\subsection{Systematic error and calibration}
We operate the balanced homodyne receiver with a local oscillator (LO) much stronger than the backscattered beam ($P_\mathrm{LO}\sim 1~\mathrm{mW}$, $P_\mathrm{bs}\sim 2~\mathrm{\mu W}$).
Excess intensity noise in the LO can manifest itself in the measured spectra as additional imprecision noise.
One can cancel this systematic error by using identical photodiodes and balancing the optical powers in front of them.
In practice, these two conditions are never exactly met, as photodiodes have always a slightly different responsivity and the optical powers need to be unbalanced to measure quadratures at angles $\theta\ne\pi/2$.
The systematic error in the spectral imprecision depends on the power unbalancing, that is, on the angle of the measured optical quadrature \cite{safavi-naeini_squeezed_2013, chen_entanglement_2020}.

\begin{figure}[hb]
  \centering
  \includegraphics{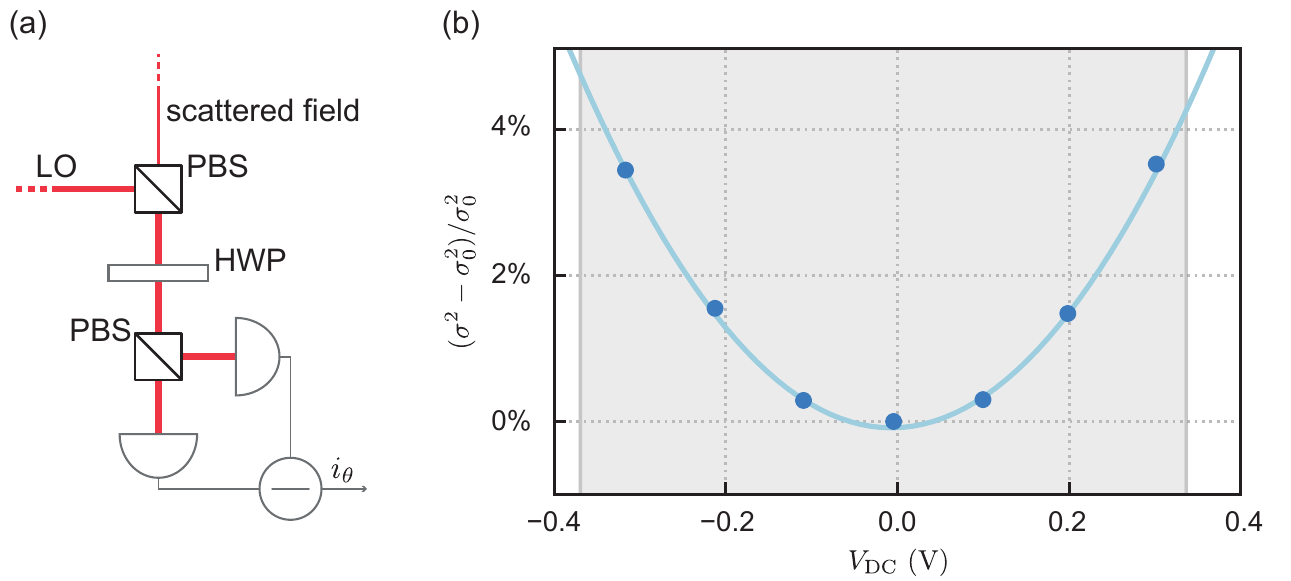}
  \caption{Homodyne receiver characterization. (a) Sketch of the homodyne receiver. The LO and the scattered field are linearly polarized along orthogonal directions. HWP, half-wave plate; PBS, polarizing beamsplitter; LO, local oscillator. (b) Relative variance (dark blue circle) as a function of the local oscillator power unbalance. The light blue line is a fit to a quadratic model. The gray shaded area corresponds to the range of unbalancing explored in the main experiment.}
  \label{fig:app_classical_noise}
\end{figure}

We characterize this error by blocking the backscattered field and unbalancing the local oscillator.
This is done by rotating the half-wave plate preceding the polarizing beam-splitter in front of the photodetector, shown in Fig.~\ref{fig:app_classical_noise}(a).
For a given position of the half-wave plate, which corresponds to a given value of power unbalancing, we record the PSD of the photodetctor output voltage.
%
Then, we integrate the PSD from $75$~kHz to $85$~kHz to obtain the variance $\sigma^2$.
In Fig.~\ref{fig:app_classical_noise}(b) we show the results for several values of the power unbalancing, which is expressed as the difference between the DC voltages of the photodiodes.
The measured variances are expressed relative to the one at the balanced configuration, $\sigma^2_0$, for which we know that the dominant source of noise is the optical shot noise \cite{tebbenjohanns_quantum_2021}.
We fit the data to a parabola (light blue), which models the measured noise in the non-ideal case \cite{safavi-naeini_squeezed_2013, chen_entanglement_2020}.
The gray shaded area indicates the region of unbalancing that we access in the experiments reported in the main text.

In the following we report the calibration procedure employed for the spectra reported in the main text.
For each angle $\theta$, we record the corresponding PSD and the photodiodes power unbalancing shown as the voltage value on the horizontal axis of Fig.~\ref{fig:app_classical_noise}(b).
We repeat this measurement for a set of angles in the range $[0, \pi]$.
Then, we block the backscattered field and we record our reference PSD corresponding to the (balanced) local oscillator only.
We compute its average spectral value around the mechanical resonance frequency and use this value to calibrate all the other PSDs, i.e., to convert them from $V^2/\text{Hz}$ to $\text{Hz}/\text{Hz}$ (shot noise units).
Finally, we subtract from each PSD the systematic error in their background, according to the measured power unbalancing and the characterization in Fig.~\ref{fig:app_classical_noise}(b).
In our regime of operation ($P_\mathrm{LO}\gg P_\text{bs}$), the excess classical noise measured by the photodiode mainly comes from the LO rather then the backscattered beam, thus it can be safely subtracted.\\
We notice that this excess noise contributes to at most $\le5\%$, much less that the maximum measured squeezing of $25\%$.
%
Thus, this systematic error does not call into question the observation of ponderomotive squeezing.

The homodyne angle $\theta$ is inferred from the DC output voltage of the balanced homodyne detector.
This voltage can be modelled as $V_\text{DC} = V_\text{off} - V_\text{amp}\cos(\theta)$.
The knowledge of the offset voltage and the peak amplitude allows us to invert this relation and express $\theta$ in terms of the measured $V_\text{DC}$.
To measure these two parameters, we sweep the angle $\theta$ by modulating the local oscillator path length with a ramp waveform and recording the oscillating output voltage $V_\text{DC}$.
The maximum and minimum recorded voltages can be promptly linked to the offset and peak amplitude values.

\subsection{Fitting the spectra}
\label{sec: fitting the spectra}

\begin{figure}[ht]
  \centering
  \includegraphics{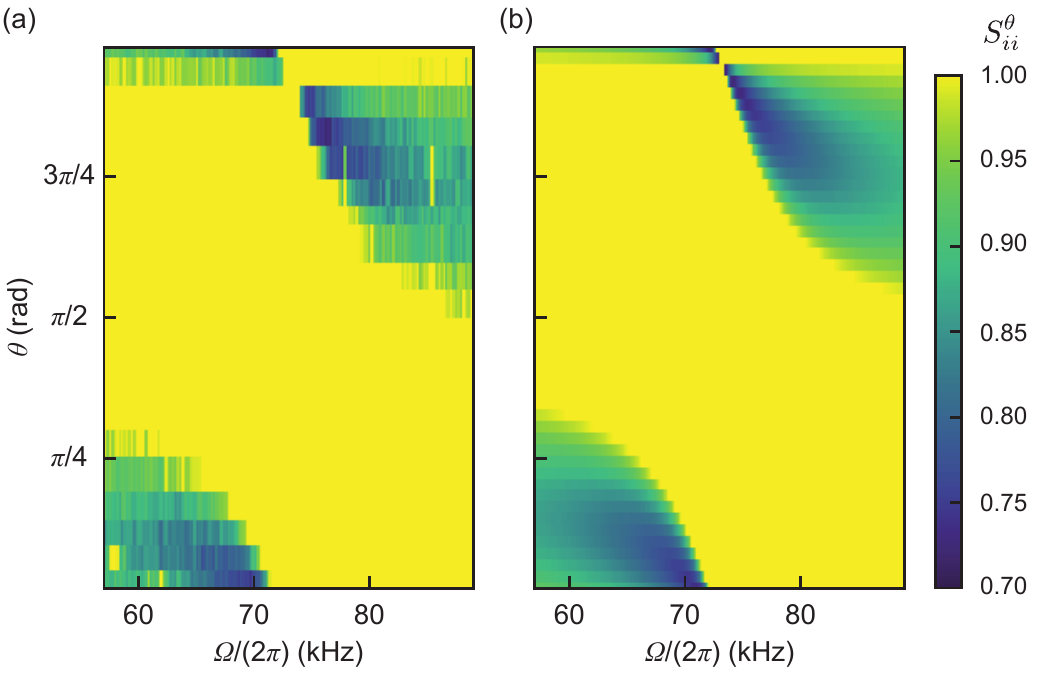}
  \caption{Homodyne spectra for different quadratures.
  (a) PSDs of the homodyne photocurrent at various angle, from $0$ to $\pi$. The spectra are normalized to the shot noise value.
  (b) Predicted spectra from the parameters extracted from the fit to Eq.~(5) of the main text.}
  \label{fig:psd_squeezing_complete}
\end{figure}

We record the PSDs at different angles in the range $[0,\pi]$, as shown in Fig.~\ref{fig:psd_squeezing_complete}(a).
We simultaneously fit all these PDSs to Eq.~(5) of the main text.
The free parameters are the mechanical damping rate $\gm$, the total decoherence rate $\Gtot$, the measurement rate $\Gmeas$, the mechanical resonance frequency $\Om$ and the angle $\theta$.
We constrain the former three parameters to be the same for all the PSDs, whereas the latter two can vary across different PSDs.
This is to take into account slow drifts of the mechanical resonance frequency across the duration of the experiment ($\sim 1~\mathrm{h}$) as well as errors in the calibration of the angle from the DC value of the detector output voltage (termed `inferred angle' from now on).

\begin{figure}[ht]
  \centering
  \includegraphics{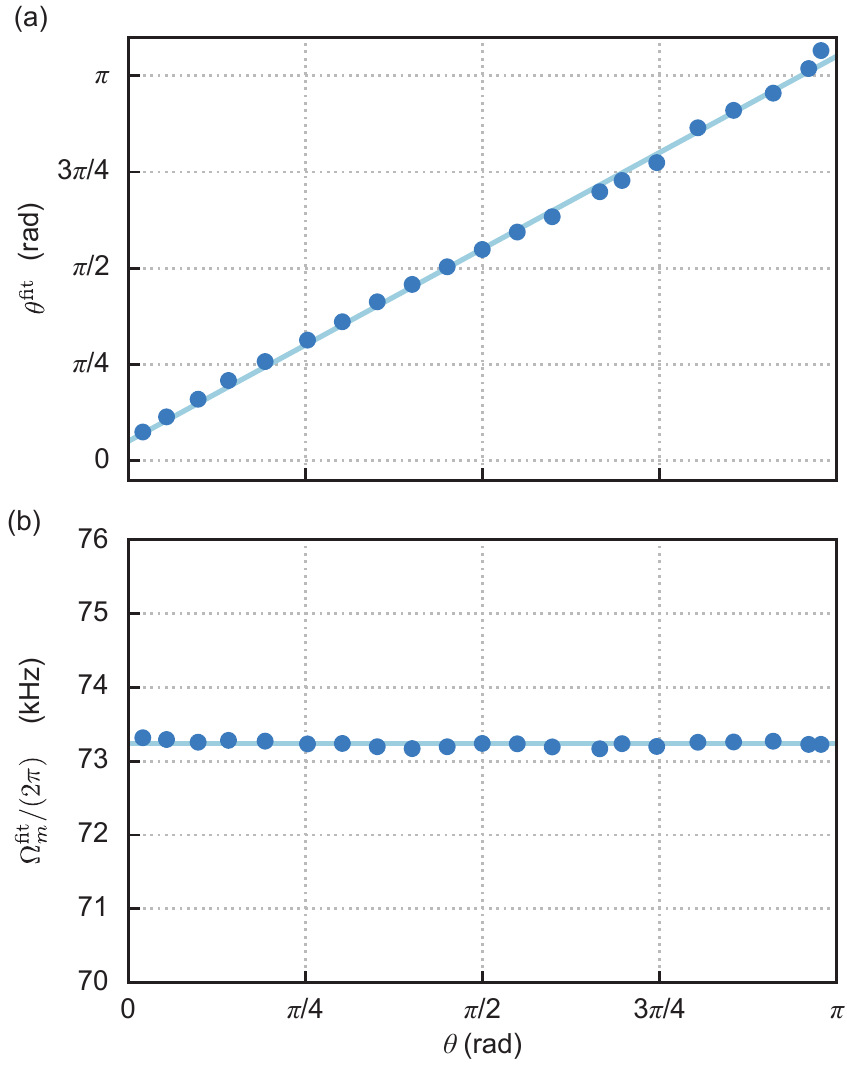}
  \caption{Results of the fits.
  (a) Fitted homodyne angle and (b) fitted mechanical resonance frequency.
  Each blue circle corresponds to a measured spectrum, labelled by the inferred angle $\theta$.
  The light blue lines in (a) and (b) show, respectively, the expected linear and constant trend.}
  \label{fig:app_fit}
\end{figure}
In Fig.~\ref{fig:app_fit}(a) and (b) we show the fitted angles and mechanical resonance frequencies for the different spectra, labelled by the inferred angle $\theta$.
We observe that the fitted angles are linearly proportional to the inferred angles, as shown by the light blue line.
The small offset of $0.05\pi$ is due to a weak field reflected by the trapping lens and co-propagating with the backscattered field, and is consistent with what was observed in Fig.~1(c) of the main text.

The mechanical resonance frequency has a mean value of $\Om/(2\pi)=73.24$~kHz and a relative standard deviation of $0.05\%$.
We hypothesize that this drift arises from slow fluctuations of the power of the trapping laser.
We use the mean value of the resonance frequency and the fitted damping rate, measurement rate and total decoherence rate to plot the predicted spectra, shown in Fig.~\ref{fig:psd_squeezing_complete}(b). 

\subsection{Wigner quasidistribution reconstruction}
\label{sec: Wigner reconstruction}

\begin{figure}[ht!]
    \centering
    \includegraphics{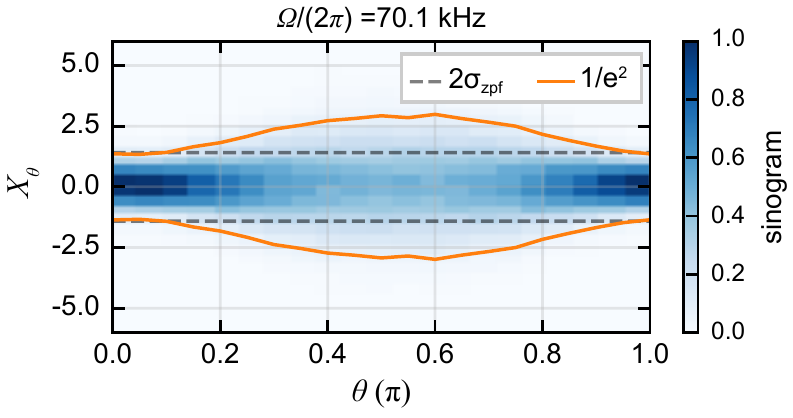}
    \caption{Sinogram of the optical mode at frequency $f=\SI{70.1}{kHz}$. Each column corresponds to a histogram of the Fourier components at frequency $f$. The dashed gray (dash-dotted orange) line corresponds to the vacuum fluctuations (variance of the histogram).}
    \label{fig: sinogram}
\end{figure}

To reconstruct the distributions in Fig.~3 of the main text, we record a time trace of the nanoparticle for values of the analyser angle $\theta$ spanning the interval $[0, \pi]$. We split the time traces in chunks of $\approx \SI{8}{ms}$ \cite{tebbenjohanns_quantum_2021}, apply a Hann window and perform a Fourier transform of each chunk. The real and imaginary parts of the Fourier transform at frequency $f$ are treated equally, in other words they are independent measurements of $X_\theta(f)$. For every frequency of interest $f$, we evaluate the Fourier transform of each chunk at $f$, concatenate real and imaginary parts and perform a histogram of the data. The histograms of the quadrature measurements as a function of the analyser angle $\theta$ constitute a so called sinogram, shown in Fig.~\ref{fig: sinogram} for $f=\SI{70.1}{kHz}$.
We then perform a numerical inverse Radon transform of the sinogram based on simultaneous algebraic reconstruction \cite{andersen_simultaneous_1984, van_der_walt_scikit-image_2014}. The covariance circle of the vacuum fluctuations, which is also used to calibrate the axes of the graphs, is estimated by blocking the particle signal, recording the shot noise of the local oscillator and estimating the variance of the signal. To estimate the covariance ellipse of the particle signal, instead, we extract three vertical cuts of the sinogram, respectively at angles $\theta_1 = 0$, $\theta_2 = \pi/4$ and $\theta_3 = \pi/2$. We fit these cuts with a gaussian distribution and extract the three respective variances $V_1, V_2, V_3$. The covariance matrix of the gaussian Wigner distribution follows from $\langle X^2 \rangle = V_1$, $\langle Y^2 \rangle = V_3$ and $\langle XY \rangle = (V_1 + V_3)/2 - V_2$.

\bibliographystyle{apsrev4-1}
\bibliography{references_MR}